\documentclass[a4paper]{jpconf}
\usepackage{graphicx}
\usepackage{amsmath}
\usepackage{amssymb}
\usepackage{rotating}
\usepackage{color}
\usepackage{multirow}
\begin{document}
\title{Dark matter in a SUSY left-right model}

\author{A.~Vicente}

\address{Laboratoire de Physique Th\'eorique, CNRS -- UMR 8627, Universit\'e de Paris-Sud 11, F-91405 Orsay Cedex, France}

\ead{Avelino.Vicente@th.u-psud.fr}

\begin{abstract}
Supersymmetric left-right models are well motivated extensions of the
Minimal Supersymmetric Standard Model since they automatically contain
the ingredients to explain the observed neutrino masses and
mixings. Here we study a SUSY model in which the left-right symmetry
is broken by triplets at a high scale leading to automatic R-parity
conservation at low energies.  The sparticle spectra in this model
differ from the usual constrained MSSM expectations and these changes
affect the relic abundance of the lightest neutralino. We discuss two
examples: (1) the standard stau co-annihilation region, and (2)
flavoured co-annihilation.
\end{abstract}

\section{Introduction}

There are good theoretical and phenomenological reasons to consider
left-right (LR) extensions
\cite{Pati:1974yy,Mohapatra:1974gc,Senjanovic:1975rk} of the
MSSM. Besides the restoration of parity at high energies, two
motivations can be especially highlighted. First, LR models contain
all the ingredients to explain the observed neutrino masses and
mixings thanks to the well-known seesaw mechanism
\cite{Minkowski:1977sc,MohSen,Schechter:1980gr,Cheng:1980qt}. And
second, the gauging of $B-L$ provides a natural framework to explain
the origin of R-parity, which can be understood as the low energy remnant
of the high energy gauge symmetry.

Here we will concentrate on the LR model proposed in
\cite{Aulakh:1997ba,Aulakh:1997fq}.  The main difference with respect
to previous proposals \cite{MohSen,Mohapatra:1980yp,Cvetic:1983su} was
the presence of additional triplets with zero $B-L$ charge. The main
advantage of this setup, in the following called the $\Omega$LR model,
is the existence of R-parity conserving minima in the tree level
scalar potential.

In the $\Omega$LR model \cite{Aulakh:1997ba,Aulakh:1997fq} new
superfields appear with masses below the GUT scale. Furthermore, the
gauge group is extended to $SU(3)_C\times SU(2)_L \times SU(2)_R
\times U(1)_{B-L}$ at energies above the LR breaking scale. Both facts
significantly affect the RGE running of all parameters.  Therefore,
even though the LR breaking scale is very far from the reach of
current accelerators, there are interesting effects in the
phenomenology caused by the running of the soft
supersymmetry breaking parameters and the subsequent deformation of the
SUSY spectra with respect to the CMSSM expectation \cite{Esteves:2010si,arXiv:1109.6478}.

Astrophysical observations and the data from WMAP
\cite{Komatsu:2010fb} put on solid grounds the existence of
non-baryonic dark matter in the universe. The most popular candidate
is the lightest neutralino in R-parity conserving
supersymmetry. However, all solutions in the CMSSM parameter space
\cite{Drees:1992am} require a fine-tuning \cite{Griest:1990kh} among
some masses in order to enhance the annihilation/co-annihilation
cross-sections, decrease the lightest neutralino relic abundance,
$\Omega_{\tilde \chi^0_1}h^2$, and reproduce the observed dark matter
relic density \cite{Komatsu:2010fb}. Therefore, it is not surprising
that even small changes in the SUSY spectra can dramatically change
the resulting $\Omega_{\tilde \chi^0_1}h^2$. The description of these
changes in the $\Omega$LR model is the main subject of this work.

\section{The model}

In this section we present the model originally defined in
\cite{Aulakh:1997ba,Aulakh:1997fq}. For further details 
see \cite{Esteves:2010si,arXiv:1109.6478}.

Below the GUT scale the gauge group of the model is $SU(3)_c \times
SU(2)_L \times SU(2)_R \times U(1)_{B-L}$. Besides the particle
content of the MSSM with the addition of (three) right-handed
neutrino(s) $\nu^c$, some additional superfields are
introduced. First, two generations of $\Phi$ superfields, bidoublets
under $SU(2)_L \times SU(2)_R$, are introduced. Furthermore, triplets
under (one of) the $SU(2)$ gauge groups are added with gauge quantum
numbers $\Delta(1,3,1,2)$, $\Delta^c(1,1,3,-2)$,
$\bar{\Delta}(1,3,1,-2)$, $\bar{\Delta}^c(1,1,3,2)$, $\Omega(1,3,1,0)$
and $\Omega^c(1,1,3,0)$.

With these representations and assuming parity conservation, the most
general superpotential compatible with the symmetries is
\begin{eqnarray} \label{eq:Wsuppot1}
{\cal W} &=& Y_Q Q \Phi Q^c 
          +  Y_L L \Phi L^c 
          - \frac{\mu}{2} \Phi \Phi
          +  f L \Delta L
          +  f^* L^c \Delta^c L^c
          + a \Delta \Omega \bar{\Delta}
          +  a^* \Delta^c \Omega^c \bar{\Delta}^c \nonumber \\
         &+& \alpha \Omega \Phi \Phi
          +  \alpha^* \Omega^c \Phi \Phi
          + M_\Delta \Delta \bar{\Delta}
          +  M_\Delta^* \Delta^c \bar{\Delta}^c
          +  M_\Omega \Omega \Omega
          +  M_\Omega^* \Omega^c \Omega^c \thickspace.
\end{eqnarray}

The breaking of the LR gauge group to the MSSM gauge group takes place
in two steps: $SU(2)_R \times U(1)_{B-L} \rightarrow U(1)_R \times
U(1)_{B-L} \rightarrow U(1)_Y$. First, the neutral component of the
$\Omega^c$ triplet takes a VEV (vacuum expectation value), $\langle
\Omega^{c \: 0} \rangle = \frac{v_R}{\sqrt{2}}$, breaking
$SU(2)_R$. Next, the group $U(1)_R \times U(1)_{B-L}$ is broken by
$\langle \Delta^{c \: 0} \rangle = \langle \bar{\Delta}^{c \: 0}
\rangle = \frac{v_{BL}}{\sqrt{2}}$. Neutrino masses are generated at
this stage via the superpotential term $f^* L^c \Delta^c L^c$
which, after $U(1)_{B-L}$ gets broken, leads to Majorana masses for
the right-handed neutrinos and a type-I seesaw mechanism.

\section{Low energy spectrum and dark matter}

In general, with $v_{BL}\le v_R < m_{GUT}$, the $\Omega$LR has a
lighter spectrum than the CMSSM for the same parameters. Since in the
$\Omega$LR model the running of the gauge couplings is changed with
respect to the MSSM case, also gaugino masses are changed. Consider
for example $M_1$. At 1-loop leading-log order we find
\begin{equation}\label{eq:M1}
M_1(m_{SUSY}) = M_{1/2} \: \left[ X_1 + X_2 \left( -3 l_1 + l_2 \right) \right]
\end{equation}
where $l_1 = \ln \left(\frac{m_{GUT}}{v_R}\right)$, $l_2 = \ln
\left(\frac{v_R}{v_{BL}}\right)$ and $X_{1,2} > 0$ are two numerical
factors.  Equation \eqref{eq:M1} shows that one can decrease
$M_1(m_{SUSY})$ with respect to the CMSSM value by using a large $l_1$
and $l_2 = 0$ and increase it only in the case $l_2 > 3 l_1$. This change
in $M_1(m_{SUSY})$ has a strong impact on dark matter phenomenology, since
the lightest neutralino is mostly bino in most parts of parameter space.

We present now our results concerning dark matter phenomenology in the
$\Omega$LR model.  We will discuss the main differences with respect
to the CMSSM for (1) the standard stau co-annihilation region
\cite{Ellis:1998kh}, and (2) flavoured co-annihilation
\cite{Choudhury:2011um}. The numerical evaluation of the parameters at
the SUSY scale was done with {\tt SPheno}
\cite{Porod:2003um,Porod:2011nf,Staub:2011dp}, using the complete
2-loop RGEs at all scales. These were derived with the Mathematica
package {\tt SARAH}
\cite{Staub:2008uz,Staub:2009bi,Staub:2010jh}. Finally, we used {\tt
  micrOmegas} \cite{Belanger:2006is} to obtain the value of
$\Omega_{\tilde\chi^0_1}h^2$

\subsection{Stau coannihilation}

As discussed, lowering the values of $v_{BL}$ and $v_R$ leads in
general to a lighter spectrum compared to the pure CMSSM case. Since
the mass of the bino decreases faster than the mass of the stau, the
quantity $\Delta_m= m_{\tilde\tau_1}-m_{\chi^0_1}$, and thus the
neutralino relic abundance, gets increased for low $v_{BL}$ and
$v_R$. This can only be compensated by reducing $m_0$ \cite{arXiv:1109.6478}.

\begin{figure}
\begin{center}
\vspace{5mm}
\includegraphics[width=0.45\textwidth]{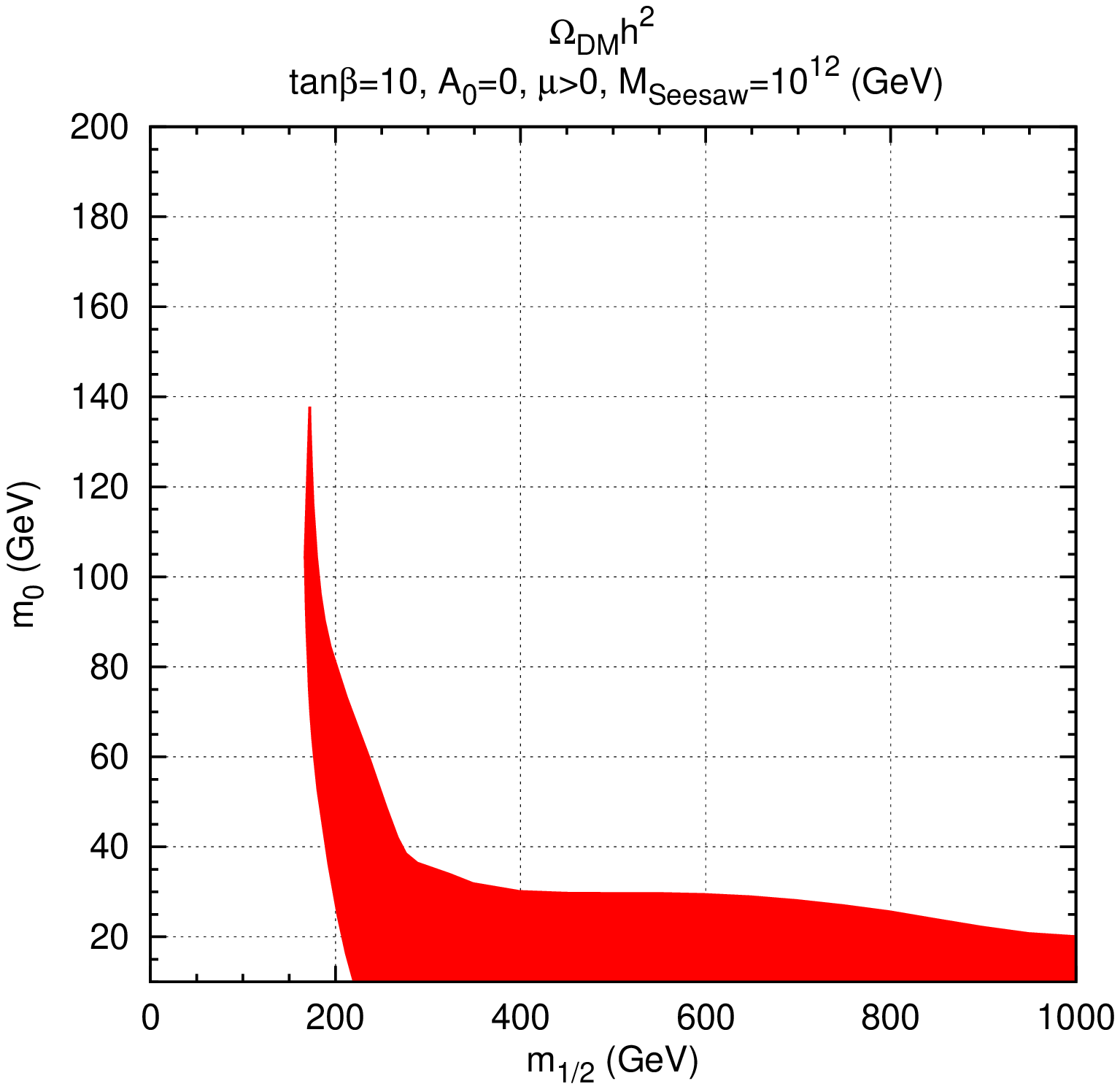}
\hspace{5mm}
\includegraphics[width=0.45\textwidth]{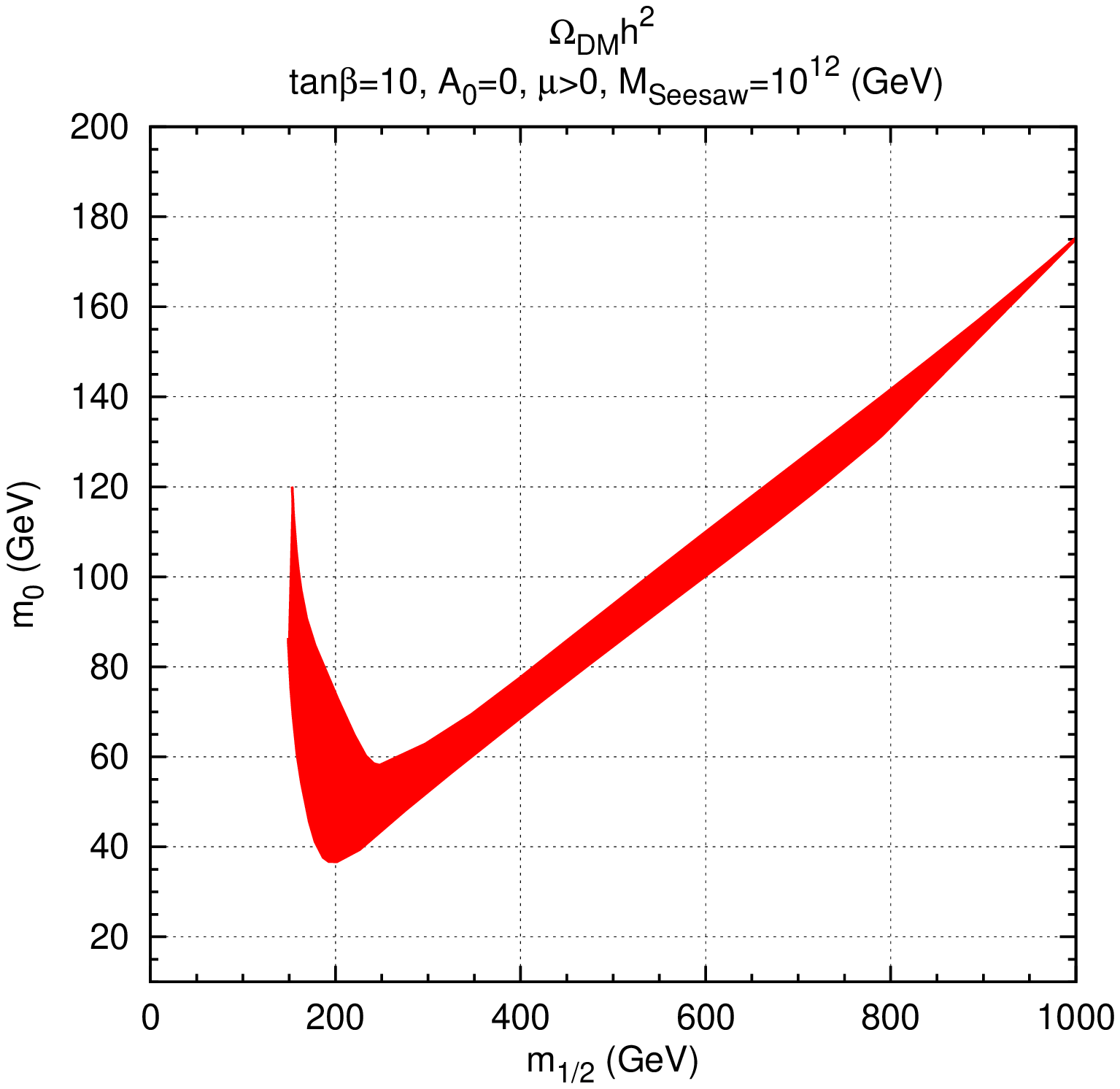}
\caption{Dark Matter 3 $\sigma$ allowed regions for $\tan\beta=10$, and 
$v_{BL}=v_R=1.5\times10^{15}$ GeV (left) and $v_{BL}=v_R=10^{16}$ GeV (right).}
\label{fig:DMtb10-I}
\end{center}
\end{figure}

Two examples of stau co-annihilation regions are shown in
Figure~\ref{fig:DMtb10-I}. As expected, a shift of the allowed region
towards smaller values of $m_0$ is found. Morever, the region can
disappear for sufficiently low $v_{BL}$ and $v_R$. The observation of
a particle spectrum consistent with the CMSSM stau co-annihilation
region could thus be turned into lower limits on $v_{BL}$ and $v_R$.

\subsection{Flavoured coannihilation}

As shown in \cite{Esteves:2010si}, the $\Omega$LR model allows for
large flavour violating effects in the right slepton sector. This
implies that the $\tilde{\tau}_R \simeq \tilde{\tau}_1$ can be made
ligther by flavour contributions, leading to co-annihilation with the
lightest neutralino in points of parameter space where it would be
impossible without flavour effects. Moreover, flavour violating
processes become relevant in the determination of the relic density,
and they must be taken into account. This is the so-called flavoured
co-annihilation mechanism \cite{Choudhury:2011um}.

\begin{figure}
\begin{center}
\vspace{5mm}
\includegraphics[width=0.45\textwidth]{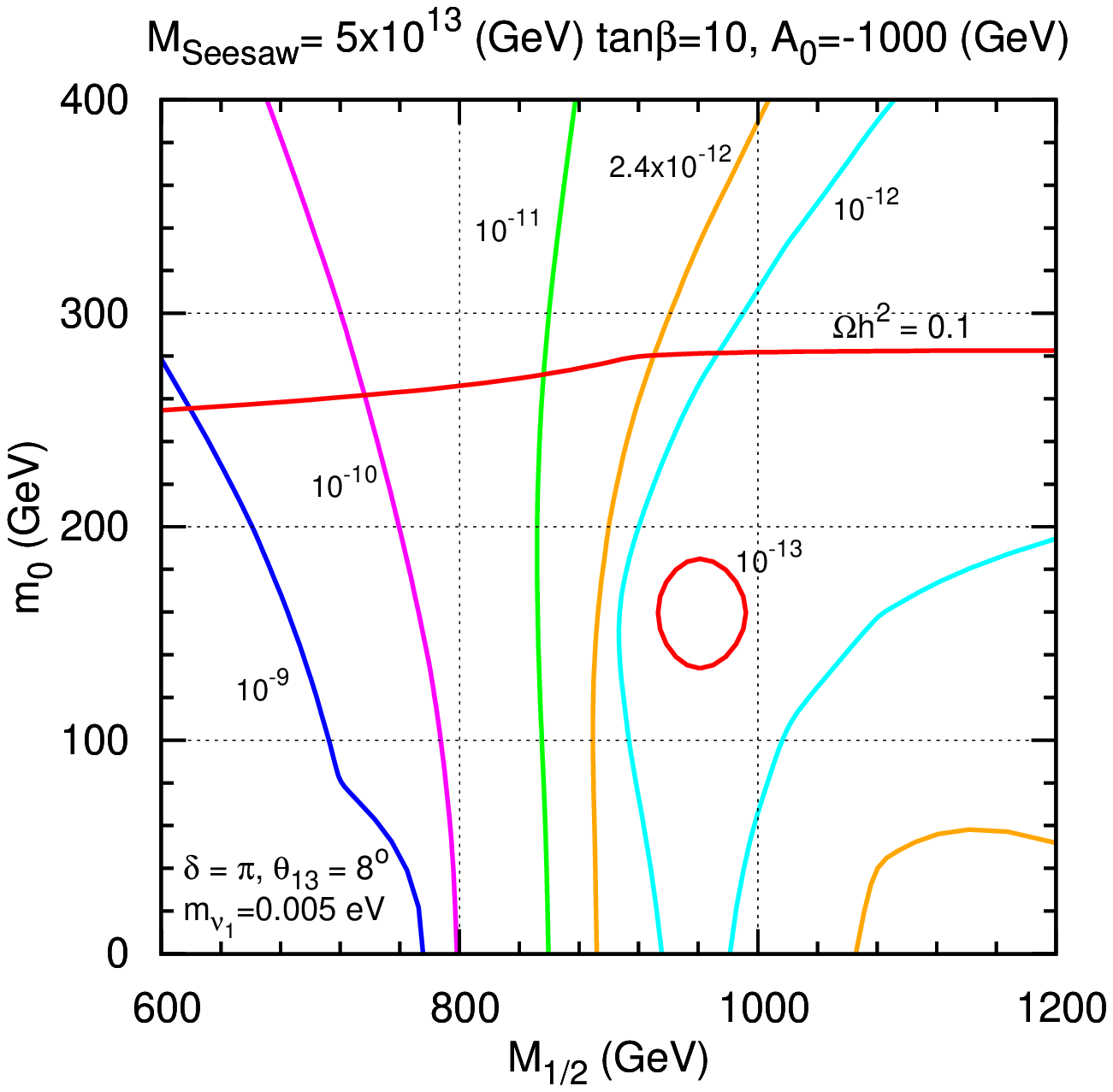}
\includegraphics[width=0.45\textwidth]{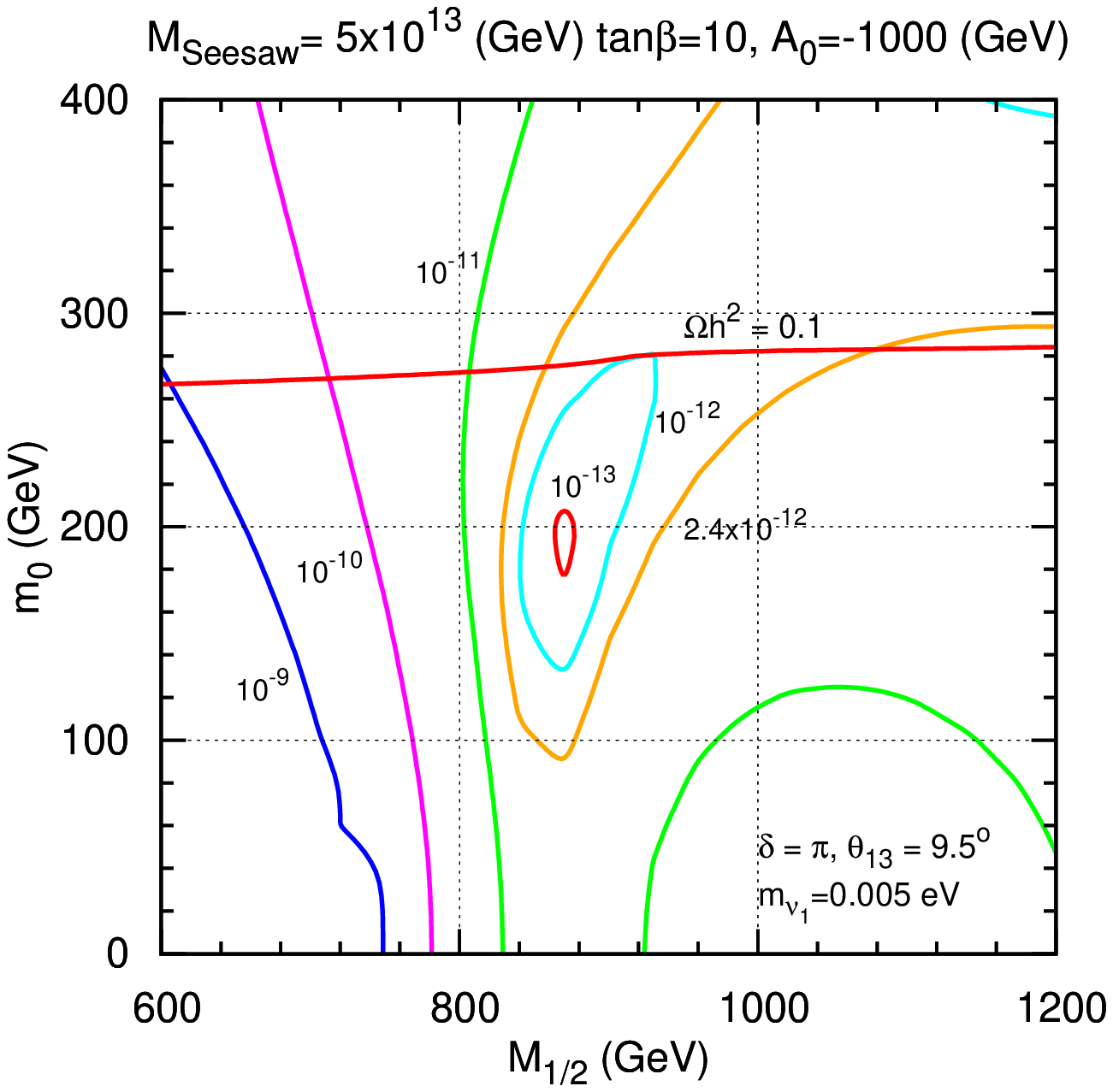}
\end{center}
\vspace{-5mm}
\caption{$\Omega_{\tilde\chi^0_1}h^2$ and $\text{Br}(\mu \to e
  \gamma)$ contour plots in the $m_0$-$M_{1/2}$ plane for
  $\delta=\pi$, $\theta_{13}=8^{0}$ (left) and $\theta_{13}=9.5^{0}$
  (right). Both examples were obtained with $v_R=10^{15}$ GeV and
  $v_{BL}=10^{14} $ GeV.}
\label{fig:flav-coa-1}
\end{figure}

When searching for such points in parameter space one must take into
account the strong limits for $\text{Br}(\mu \to e \gamma)$ given by
the MEG experiment \cite{Adam:2011ch}. A fine-tuning of the parameters
is required in order to cancel this observable and, at the same time,
reduce the $\tilde{\tau}_R$ mass
sufficiently. Figure~\ref{fig:flav-coa-1} shows two examples. On the
left panel the cancellation of $\text{Br}(\mu \to e \gamma)$ is
obtained with $\delta = \pi$ and $\theta_{13} = 8^{\circ}$, while on
the right panel with $\theta_{13} = 9.5^{\circ}$. Note that this
occurs for different values of $m_0$ and $M_{1/2}$.

\section{Conclusions}

We have studied the dark matter phenomenology of the $\Omega$LR model,
a supersymmetric left-right model that breaks the LR symmetry at high
energies. The change in the running of the parameters at energies
above the breaking scale has a strong impact on the low energy
spectra, modifying the usual dark matter predictions of the CMSSM.

The shift of the stau co-annihilation region and its potential
disappearance has been studied. The fact that the lightest neutralino
is usually lighter in the $\Omega$LR model than in the CMSSM reduces
the allowed stau co-annihilation region, which can even vanish for low
$v_R$ and/or $v_{BL}$.

The model also allows for flavoured co-annihilation. Examples have been
given of points in parameter space where the $\tilde{\tau}_R$ mass is
reduced by means of flavour effects, leading to the observed dark matter
relic abundance.

\section*{Acknowledgements}

This work has been done in collaboration with J. N. Esteves,
M. Hirsch, W. Porod, J. C. Rom\~ao and F. Staub. A.V. acknowledges
support from the ANR project CPV-LFV-LHC {NT09-508531}. A.V. was also
partially suported by Marie Curie Early Initial Training Network
Fellowships of the European Community's Seventh Framework Programme
under contract number (PITN-GA-2008-237920-UNILHC) and by the DFG
project number PO-1337/1-1.


\section*{References}

\begin{thebibliography}{9}

\bibitem{Pati:1974yy}
  J.~C.~Pati, A.~Salam,
  Phys.\ Rev.\  D {\bf 10} (1974) 275 
  [Erratum-ibid.\  D {\bf 11} (1975) 703].

\bibitem{Mohapatra:1974gc}
  R.~N.~Mohapatra, J.~C.~Pati,
  Phys.\ Rev.\  D {\bf 11} (1975) 2558.

\bibitem{Senjanovic:1975rk}
  G.~Senjanovic, R.~N.~Mohapatra,
  Phys.\ Rev.\  D {\bf 12} (1975) 1502.

\bibitem{Minkowski:1977sc}
  P.~Minkowski,
  Phys.\ Lett.\ B {\bf 67} (1977) 421.
T.~Yanagida, in {\it KEK lectures}, ed.\  O.~Sawada and A.~Sugamoto,
KEK, 1979;
M Gell-Mann, P Ramond, R. Slansky, in {\it Supergravity}, ed.\ P.\ van
Niewenhuizen and D.\ Freedman (North Holland, 1979);

\bibitem{MohSen}
R.N.~Mohapatra, G.~Senjanovic, {\sl Phys. Rev. Lett.}\/ {\bf 44}
 (1980) 912.

\bibitem{Schechter:1980gr}
  J.~Schechter, J.~W.~F.~Valle,
  Phys.\ Rev.\ D {\bf 22}  (1980) 2227.

\bibitem{Cheng:1980qt}
  T.~P.~Cheng, L.~F.~Li,
  Phys.\ Rev.\  D {\bf 22} (1980) 2860.

\bibitem{Aulakh:1997ba}
  C.~S.~Aulakh, K.~Benakli, G.~Senjanovic,
  Phys.\ Rev.\ Lett.\  {\bf 79} (1997) 2188 
  [arXiv:hep-ph/9703434].

\bibitem{Aulakh:1997fq}
  C.~S.~Aulakh, A.~Melfo, A.~Rasin, G.~Senjanovic,
  Phys.\ Rev.\  D {\bf 58}  (1998) 115007
  [arXiv:hep-ph/9712551].

\bibitem{Mohapatra:1980yp}
R.~N.~Mohapatra, G.~Senjanovic,
Phys.\ Rev.\  {\bf D23} (1981) 165.
  
\bibitem{Cvetic:1983su}
  M.~Cvetic, J.~C.~Pati,
  Phys.\ Lett.\  B {\bf 135} (1984) 57.

\bibitem{Esteves:2010si}
J.~N.~Esteves, J.~C.~Romao, M.~Hirsch, A.~Vicente, W.~Porod, F.~Staub,
JHEP {\bf 1012 } (2010)  077
[arXiv:1011.0348 [hep-ph]].

\bibitem{arXiv:1109.6478}
  J.~N.~Esteves, J.~C.~Romao, M.~Hirsch, W.~Porod, F.~Staub and A.~Vicente,
  arXiv:1109.6478 [hep-ph].

\bibitem{Komatsu:2010fb}
  E.~Komatsu {\it et al.},
  arXiv:1001.4538 [astro-ph.CO].

\bibitem{Drees:1992am}
  M.~Drees, M.~M.~Nojiri,
  Phys.\ Rev.\  D {\bf 47}  (1993) 376
  [arXiv:hep-ph/9207234].

\bibitem{Griest:1990kh}
  K.~Griest and D.~Seckel,
  Phys.\ Rev.\  D {\bf 43}  (1991) 3191.

\bibitem{Ellis:1998kh}
  J.~R.~Ellis, T.~Falk, K.~A.~Olive,
  Phys.\ Lett.\  B {\bf 444} (1998) 367
  [arXiv:hep-ph/9810360].

\bibitem{Choudhury:2011um}
D.~Choudhury, R.~Garani, S.~K.~Vempati,
arXiv:1104.4467 [hep-ph].

\bibitem{Porod:2003um}
  W.~Porod,
  Comput.\ Phys.\ Commun.\  {\bf 153} (2003) 275 
  [arXiv:hep-ph/0301101].

\bibitem{Porod:2011nf}
  W.~Porod, F.~Staub,
  arXiv:1104.1573 [hep-ph].

\bibitem{Staub:2011dp}
  F.~Staub, T.~Ohl, W.~Porod, C.~Speckner,
  arXiv:1109.5147 [hep-ph].

\bibitem{Staub:2008uz}
  F.~Staub,
  arXiv:0806.0538 [hep-ph].

\bibitem{Staub:2009bi}
  F.~Staub,
  Comput.\ Phys.\ Commun.\  {\bf 181}  (2010) 1077
  [arXiv:0909.2863 [hep-ph]].

\bibitem{Staub:2010jh}
  F.~Staub,
  Comput.\ Phys.\ Commun.\  {\bf 182 } (2011)  808
  [arXiv:1002.0840 [hep-ph]].

\bibitem{Adam:2011ch}
  J.~Adam {\it et al.} [MEG Collaboration],
  arXiv:1107.5547 [hep-ex].

\bibitem{Belanger:2006is}
  G.~Belanger, F.~Boudjema, A.~Pukhov, A.~Semenov,
  Comput.\ Phys.\ Commun.\  {\bf 176}  (2007) 367
  [arXiv:hep-ph/0607059].

\end{thebibliography}
\end{document}